\documentclass[aps,prc,reprint,superscriptaddress,amsmath,amssymb,nofootinbib]{revtex4-2}

\usepackage{setspace}
\usepackage{graphicx}
\usepackage{bm}
\usepackage[utf8]{inputenc}
\usepackage[colorlinks=true,allcolors=blue]{hyperref}
\usepackage{tabularx}
\usepackage{aasmacros}
\usepackage[normalem]{ulem}

\begin{document}

\title{Revised $^{45}$V($p,\gamma$)$^{46}$Cr reaction rate and its impact on the production of $^{44}$Ti in core-collapse supernovae}

\author{R.~S.~Sidhu}
\email{ragan.sidhu@surrey.ac.uk}
\affiliation{School of Mathematics and Physics, University of Surrey, Guildford, GU2 7XH, United Kingdom}

\author{Y.~Luo}
\email{ydongluo@ibs.re.kr}
\affiliation{Center for Exotic Nuclear Studies, Institute for Basic Science, 55, Expo-ro, Yuseong-gu, Daejeon, Korea, 34126}

\author{C.~Sarma}
\affiliation{School of Mathematics and Physics, University of Surrey, Guildford, GU2 7XH, United Kingdom}

\author{M.~Wiescher}
\affiliation{Department of Physics and Astronomy, University of Notre Dame, Notre Dame, Indiana 46556, USA}

\author{X.~Xu}
\affiliation{Institute of Modern Physics, Chinese Academy of Sciences, Lanzhou 730000, China}

\date{\today}

\begin{abstract}

The thermonuclear $^{45}$V($p,\gamma$)$^{46}$Cr reaction is the primary leakage pathway from the $^{44}$Ti--$^{45}$V quasi-equilibrium cluster during $\alpha$-rich freeze-out in core-collapse supernova (CCSN), and therefore plays a key role in determining the final abundance of the $\gamma$-ray-emitting isotope $^{44}$Ti. 
A recent high-resolution $\gamma$-ray spectroscopy study [C. Cousins \textit{et al.}, \href{https://doi.org/10.1103/9zv2-wlkl}{Phys. Rev. Lett. 136, 252701 (2026)}] identified ten previously unknown low-spin proton-unbound states in $^{46}$Cr, enabling the first experimentally constrained $^{45}$V($p,\gamma$)$^{46}$Cr reaction rate. 
That study adopted the AME2020 mass excess of $^{46}$Cr, ME($^{46}$Cr) = $-29472(11)$ keV. In the present work, we use the more precise mass excess ME($^{46}$Cr) = $-29477.2(2.6)$ keV measured using the CSRe storage ring [M.~Wang \textit{et al.}, \href{https://doi.org/10.1103/PhysRevC.106.L051301}{Phys. Rev. C \textbf{106}, L051301 (2022)}], which improves the mass precision by more than a factor of four.
We recalculate the $^{45}$V($p,\gamma$)$^{46}$Cr reaction rate using this improved mass value. 
In addition to proton capture on the ground state of $^{45}$V, we also include proton capture on the first and second excited states of $^{45}$V and perform new shell-model calculations of the corresponding proton spectroscopic factors.
The precise $^{46}$Cr mass reduces the mass-related contribution to the reaction-rate uncertainty to a subdominant level, with the remaining uncertainty dominated by the proton spectroscopic factors and $\gamma$-decay widths of the resonant states in $^{46}$Cr.
The revised reaction rate is up to 69\% higher than that of Cousins \textit{et al.} over the temperature range relevant to $\alpha$-rich freeze-out $T \simeq$ 1.5--2~GK. 
We further investigate the impact of the revised reaction rate on $^{44}$Ti production using CCSN nucleosynthesis calculations. 
Relative to the rate of The \textit{et al.} [\href{https://iopscience.iop.org/article/10.1086/306057}{ApJ \textbf{504}, 500 (1998)}], the revised rate increases the ejected $^{44}$Ti yield by  $\sim$26\% in a $20\,M_\odot$ model, but produces a negligible change for the SN~1987A trajectory. We show that the sensitivity of $^{44}$Ti production to the $^{45}$V($p,\gamma$)$^{46}$Cr reaction is governed by the electron fraction ($Y_e$) of the $^{44}$Ti-producing ejecta: the reaction has a significant impact in proton-rich ejecta with $Y_e \approx 0.50$, but little effect in more neutron-rich ejecta with $Y_e \approx 0.496$, where the lower free-proton abundance suppresses the reaction flow. These results reconcile the apparently conflicting conclusions of previous sensitivity studies.
\end{abstract}


\maketitle
 
\section{Introduction}

Core-collapse supernovae (CCSNe) are among the most energetic events in the Universe and play a central role in the synthesis and dispersal of heavy elements~\cite{bethe1990supernova}. 
Among the radioactive nuclei produced in these explosions, $^{44}$Ti ($t_{1/2}=59.1(3)$ y) is a key diagnostic of explosive nucleosynthesis because its lifetime is long compared with the explosion timescale but sufficiently short to allow direct observation in young supernova remnants. 
The characteristic $\gamma$ rays at 67.9, 78.4, and 1157.0~keV, emitted in the decay chain
$^{44}$Ti$\rightarrow$$^{44}$Sc$\rightarrow$$^{44}$Ca, have been detected in Cassiopeia A (Cas A)~\cite{Iyudin1994,Vink2001,Grefenstette2014} and Supernova 1987A (SN1987A)~\cite{grebenev2012hard,boggs201544ti} through $\gamma$-ray and X-ray observations, providing direct evidence for ongoing radioactive decay in recently synthesized ejecta. Measurements of the ejected $^{44}$Ti mass have therefore become an important observational benchmark for testing models of supernova explosions and nucleosynthesis~\cite{wang2024insights,sieverding2023production}.

The production of $^{44}$Ti has long been predicted to occur in the innermost ejecta of CCSN during explosive silicon burning and the subsequent $\alpha$-rich freeze-out from nuclear statistical equilibrium (NSE)~\cite{Woosley1973,Thielemann1996}. During the passage of the supernova shock through silicon-rich layers, the rapidly expanding and cooling material departs from NSE before $\alpha$ particles can fully reassemble into iron-group nuclei, leaving an enhanced abundance of free $\alpha$ particles that strongly influences the reaction flow~\cite{Woosley2002}. Since $^{44}$Ti is synthesized close to the boundary between material that is ejected and material that falls back onto the compact remnant, its final abundance is particularly sensitive to the explosion dynamics, including the peak temperature, entropy, expansion timescale, electron fraction ($Y_e$), and the location of the mass cut~\cite{The1998,Magkotsios2010,Subedi2020,Hermansen2020}. Consequently, observed $^{44}$Ti yields provide valuable constraints on both the explosion mechanism and the nuclear physics governing nucleosynthesis in the mass region $A\approx40$--50.

The astrophysical importance of $^{44}$Ti extends beyond direct observations of young supernova remnants. Excesses of its stable daughter nucleus, $^{44}$Ca, found in presolar grains provide complementary evidence for supernova nucleosynthesis and preserve a record of the isotopic composition of individual stellar ejecta~\cite{Amari1992,Nittler1996,The1998}. Together, astronomical observations and laboratory measurements establish $^{44}$Ti as one of the most powerful probes of explosive nucleosynthesis in CCSN.

Although the thermodynamic conditions of the explosion largely determine where $^{44}$Ti is synthesized, the final abundance also depends on the underlying nuclear reaction rates. Early reaction-network sensitivity studies identified several key reactions influencing $^{44}$Ti production, with the $^{45}$V($p,\gamma$)$^{46}$Cr reaction identified as one of the most important reactions under nearly symmetric matter conditions ($Y_e\approx0.50$)~\cite{The1998}. 
More comprehensive investigations subsequently demonstrated that $^{44}$Ti synthesis is governed not only by classical $\alpha$-rich freeze-out, but also by complex transitions between equilibrium states, reorganizations of the reaction flow, and multidimensional expansion trajectories~\cite{Magkotsios2010}. These studies introduced the concept of the $^{44}$Ti ``chasm'', a region in temperature-density space where the final $^{44}$Ti abundance decreases dramatically, and confirmed that the $^{45}$V($p,\gamma$)$^{46}$Cr reaction remains one of the dominant nuclear-physics uncertainties governing the final $^{44}$Ti abundance over a broad range of astrophysical conditions~\cite{Magkotsios2010}.

During $\alpha$-rich freeze-out, reactions linking the silicon-calcium and iron-group quasi-statistical equilibrium (QSE) clusters determine the redistribution of nuclear flow. Production of $^{44}$Ti proceeds primarily through the reactions $^{40}$Ca($\alpha,\gamma$)$^{44}$Ti and $^{40}$Ca($\alpha,p$)$^{43}$Sc($p,\gamma$)$^{44}$Ti, while destruction occurs mainly through $^{44}$Ti($\alpha,p$)$^{47}$V, $^{44}$Ti($\alpha,\gamma$)$^{48}$Cr, and $^{44}$Ti($p,\gamma$)$^{45}$V. Owing to its relatively small $Q$-value ($Q=1.627$ MeV), the $^{44}$Ti($p,\gamma$)$^{45}$V reaction remains in equilibrium with its inverse photodisintegration reaction over much of the relevant temperature range, leading to an equilibrium abundance ratio~\cite{The1998}

\begin{equation}
\frac{Y(^{45}\mathrm{V})}{Y(^{44}\mathrm{Ti})}
\propto Y_p
\exp\left(\frac{11.605\,Q}{T_9}\right),
\label{eq.1}
\end{equation}
where $Q$ is given in MeV, $T_9$ is the temperature in units of $10^9$ K, and $Y_p$ is the proton abundance. In contrast, the subsequent reaction $^{45}$V($p,\gamma$)$^{46}$Cr has a considerably larger $Q$-value ($Q=4.874$ MeV), preventing equilibrium with the reverse photodisintegration channel. 
Instead, it acts as an efficient leakage pathway that transfers material from the $^{44}$Ti QSE cluster toward heavier nuclei. The strength of this leakage therefore has a direct impact on the final abundance of $^{44}$Ti.

Despite its astrophysical importance, the $^{45}$V($p,\gamma$)$^{46}$Cr reaction rate remains poorly constrained owing to the limited experimental information available on the resonant states in $^{46}$Cr. Recently, \citet{9zv2-wlkl} identified, for the first time, ten previously unknown proton-unbound states in $^{46}$Cr using in-beam $\gamma$-ray spectroscopy at Argonne National Laboratory. By combining these measurements with mirror-state information from $^{46}$Ti, they derived the first experimentally constrained thermonuclear reaction rate for $^{45}$V($p,\gamma$)$^{46}$Cr. 
However, their calculation adopted the AME2020 mass excess of $^{46}$Cr, $\mathrm{ME}=-29472(11)$ keV~\cite{wang2021ame}, and identified the uncertainty in the corresponding proton separation energy as one of the dominant sources of uncertainty in the calculated reaction rate.

Since the AME2020 evaluation, the mass of $^{46}$Cr has been measured with substantially improved precision using the $B\rho$-defined isochronous mass spectrometry ($B\rho$-IMS) technique at the Cooler Storage Ring (CSRe) in Lanzhou, yielding a mass excess of $\mathrm{ME}=-29477.2(2.6)$ keV~\cite{wang2022b}.
This measurement represents more than a fourfold improvement in precision over the previously adopted value and corresponds to a proton separation energy of $S_p=4879.8(2.8)$ keV.

In the present work, we combine this improved mass measurement with the recently reported resonance information from~\citet{9zv2-wlkl} to derive an updated thermonuclear $^{45}$V($p,\gamma$)$^{46}$Cr reaction rate.
In addition, we include proton capture on the first and second excited states of $^{45}$V at 56.7(5) and 56.8(6) keV~\cite{gronemeyer1980gamma,bentley2006high,dossat2007decay,borrel1992decay,BURROWS2008171}, which are in thermal equilibrium under $\alpha$-rich freeze-out conditions, and perform new shell-model calculations of the corresponding proton spectroscopic factors.
To assess the astrophysical impact of the revised rate, we perform parameterized $\alpha$-rich freeze-out calculations over a range of peak temperatures, densities, and electron fractions to identify the conditions under which $^{44}$Ti production is sensitive to this reaction. We also carry out post-processing nucleosynthesis calculations using ejecta trajectories from a $20\,M_\odot$ CCSN model~\cite{2019ApJS..243...10P} and an SN~1987A progenitor model~\cite{1996ApJ...460..408T,2015PTEP.2015f3E01K}. 
These calculations demonstrate that the sensitivity of $^{44}$Ti production to the $^{45}$V($p,\gamma$)$^{46}$Cr reaction is governed primarily by the electron fraction, or equivalently the free-proton abundance, of the supernova ejecta.

\section{Thermonuclear Reaction Rate}

The total thermonuclear reaction rate of the $^{45}$V($p,\gamma$)$^{46}$Cr reaction is obtained as the incoherent sum of resonant and non-resonant direct-capture (DC) contributions from the ground state and thermally excited states of the $^{45}$V target nucleus, weighted by individual population factors~\cite{Fowler1967}

\begin{table*}
\label{tab:table1}
\begin{ruledtabular}
\caption{Properties of resonant states in $^{46}$Cr. Listed are excitation energy $E_x$, spin and parity $J^{\pi}$, centre-of-mass resonance energy $E_R$, spectroscopic factors $C^2S$, proton decay width $\Gamma_p$, $\gamma$-decay width $\Gamma_\gamma$, and the resonance strength $\omega \gamma$. The upper part is for ground-state capture; the middle part is for capture on the first excited state in $^{45}$V; the lower part is for capture on the second excited state in $^{45}$V. The $E_x$, $J^{\pi}$, and $\Gamma_\gamma$ are taken from~\citet{9zv2-wlkl}. Only the central values of $C^2S$, $\Gamma_p$, $\Gamma_\gamma$, and $\omega\gamma$ are listed; the corresponding uncertainties are given in Tables I--III in~\cite{Sidhu_supplement}. } 

\begin{tabular}{ c c c c c c c }                    
$E_x$	(keV)	&	$J^\pi$	&	$E_R$	(keV)	&	$C^2S$	&	$\Gamma_p$	(eV)	&	$\Gamma_\gamma$	(eV)	&	$\omega	\gamma$	(eV)	\\	\hline
4971(6)	&	($6^+$)	&	91.2(6.6)	&	0.0184	&	1.87E-26	&	9.51E-02	&	1.52E-26	\\							
5194(1)	&	$2^+$	&	314.2(2.9)	&	0.1617	&	8.35E-08	&	3.49E-01	&	2.61E-08	\\							
5305(5)	&	($4^+$)	&	425.2(5.7)	&	0.0222	&	3.12E-06	&	4.81E-02	&	1.75E-06	\\							
5411(1)	&	($4^+$)	&	531.2(2.9)	&	0.2481	&	1.28E-03	&	4.98E-01	&	7.18E-04	\\							
5588(2)	&	$2^+$	&	708.2(3.4)	&	0.1077	&	3.30E-02	&	7.22E-01	&	9.55E-03	\\							
5730(6)	&	($5^+$)	&	850.2(6.6)	&	0.1802	&	5.46E-01	&	9.21E-02	&	5.41E-02	\\							
5792(2)	&	$2^+$	&	912.2(3.4)	&	0.0840	&	5.81E-01	&	8.75E-01	&	7.39E-02	\\							
5829(6)	&	($3^-$)	&	949.2(6.6)	&	0.0010	&	2.43E-02	&	3.16E-01	&	9.40E-03	\\							
6665(5)	&	($2^-$)	&	1785.2(5.7)	&	0.0001	&	5.66E-02	&	3.69E-01	&	6.08E-03	\\							
7120(7)	&	($4^+$)	&	2240.2(7.5)	&	0.0517	&	1.15E+03	&	8.12E-01	&	3.98E-01	\\	\hline						
4971(6)	&	($6^+$)	&	34.5(6.6)	&	0.0328	&	1.75E-46	&	9.51E-02	&	1.89E-46	\\							
5194(1)	&	$2^+$	&	257.5(3.0)	&	0.1286	&	1.01E-09	&	3.49E-01	&	4.21E-10	\\							
5305(5)	&	($4^+$)	&	368.5(5.7)	&	0.0433	&	4.77E-07	&	4.81E-02	&	3.58E-07	\\							
5411(1)	&	($4^+$)	&	474.5(3.0)	&	0.0925	&	8.08E-05	&	4.98E-01	&	6.04E-05	\\							
5588(2)	&	$2^+$	&	651.5(3.4)	&	0.0788	&	7.85E-03	&	7.22E-01	&	3.03E-03	\\							
5730(6)	&	($5^+$)	&	793.5(6.6)	&	0.0565	&	5.06E-04	&	9.21E-02	&	6.69E-05	\\							
5792(2)	&	$2^+$	&	855.5(3.4)	&	0.0606	&	1.98E-01	&	8.75E-01	&	3.36E-02	\\							
5829(6)	&	($3^-$)	&	892.5(6.6)	&	0.0013	&	1.57E-02	&	3.16E-01	&	8.09E-03	\\							
6665(5)	&	($2^-$)	&	1728.5(5.7)	&	0.0000	&	0.00E+00	&	3.69E-01	&	0.00E+00	\\							
7120(7)	&	($4^+$)	&	2183.5(7.5)	&	0.0090	&	1.69E+02	&	8.12E-01	&	7.77E-02	\\	\hline						
4971(6)	&	($6^+$)	&	34.4(6.6)	&	0.0000	&	0.00E+00	&	9.51E-02	&	0.00E+00	\\							
5194(1)	&	$2^+$	&	257.4(3.0)	&	0.0581	&	4.53E-10	&	3.49E-01	&	2.83E-10	\\							
5305(5)	&	($4^+$)	&	368.4(5.7)	&	0.0250	&	1.27E-09	&	4.81E-02	&	1.43E-09	\\							
5411(1)	&	($4^+$)	&	474.4(3.0)	&	0.0034	&	1.52E-08	&	4.98E-01	&	1.71E-08	\\							
5588(2)	&	$2^+$	&	651.4(3.5)	&	0.1609	&	1.60E-02	&	7.22E-01	&	9.27E-03	\\							
5730(6)	&	($5^+$)	&	793.4(6.6)	&	0.0010	&	8.94E-06	&	9.21E-02	&	1.77E-06	\\							
5792(2)	&	$2^+$	&	855.4(3.5)	&	0.1519	&	4.95E-01	&	8.75E-01	&	1.26E-01	\\							
5829(6)	&	($3^-$)	&	892.4(6.6)	&	0.0023	&	1.19E-03	&	3.16E-01	&	9.23E-04	\\							
6665(5)	&	($2^-$)	&	1728.4(5.7)	&	0.0001	&	6.48E-01	&	3.69E-01	&	1.39E-01	\\							
7120(7)	&	($4^+$)	&	2183.4(7.5)	&	0.0050	&	1.83E+00	&	8.12E-01	&	1.26E-03	\\							

\end{tabular}
\label{tab:table1}
\end{ruledtabular}
\end{table*}

\begin{multline}
    N_A \langle\sigma v\rangle_{\rm{total}} = \sum_{i} ( N_A  \langle\sigma v\rangle_{\mathrm{R},i} + \; N_A  \langle\sigma v\rangle_{\mathrm{NR},i}) \\ \times \frac{(2J_i + 1 )e^{-E_i/kT}}{\sum_n (2J_n+1)e^{-E_n/kT}} ,
\label{eq.2}
\end{multline}
where $i$ represents the initial state of the $^{45}$V target nucleus. In the present case, the included states are the ground state ($E=0$ keV, $J^\pi=7/2^-$~\cite{BURROWS2008171}), the first excited state ($E=56.7(5)$ keV, $J^\pi=5/2^-$~\cite{BURROWS2008171}), and the second excited state ($E=56.8(6)$ keV, $J^\pi=3/2^-$~\cite{BURROWS2008171}) in $^{45}$V. $n$ corresponds to these three thermally populated states.

For isolated narrow resonances, the resonant contribution for proton capture from an initial state $i$ of $^{45}$V is obtained by summing over all ten proton unbound resonance states $j$ in the compound nucleus $^{46}$Cr, and is given by~\cite{Fowler1967,rolfs1990r}

\begin{multline}
N_A \langle\sigma v\rangle_{{\mathrm{R},i}} \left[\frac{\rm{cm^3}}{\rm s\cdot mol}\right] = \frac{1.5394 \times 10^{11}} {(\mu T_{9})^{3/2}}  \times \sum_{j} (\omega \gamma_{ij} \times 10^{-6}) \\  \times \mathrm{exp}\left( \frac{-11.6045 \times E_{{\rm{R}},ij}}{1000 \times T_{9}}\right), 
\label{eq.3}
\end{multline}
where $N_A$ is Avogadro’s number, $\mu$ is the reduced mass in atomic mass units, $T_9$ is the temperature in GK, $\omega\gamma_{ij}$ is the resonance strength in eV, and $E_{\mathrm{R},ij}$ is the centre-of-mass resonance energy (keV), defined as $E_{\mathrm{R},ij} = E_j - S_p(^{46}$Cr) $-E_i$, 
with $E_j$ the excitation energy of the resonant state in $^{46}$Cr and 
$E_i$ is the initial state energy of the $^{45}$V. In the present work, we use a proton separation energy of $S_p(^{46}$Cr) = 4879.8(2.8) keV, calculated using the the mass excesses ME($^{45}$V) = $-31886.4 \pm 0.9$ keV~\cite{wang2021ame} (AME2020) and ME($^{46}$Cr) = $-29477.2 \pm 2.6$ keV~\cite{wang2022b}. This value is 5.8 keV higher than the proton separation energy of 4874(11) keV adopted by~\citet{9zv2-wlkl} and has an uncertainty that is smaller by a factor of 3.9.

The resonance strength is calculated using the Breit--Wigner formalism for isolated resonances~\cite{BreitWigner1936}
\begin{equation}
   \omega \gamma_{ij} = \frac{2J_j + 1}{(2J_p + 1)(2J_i + 1)}\frac{\Gamma_{p, ij} \Gamma_{\gamma,j}}{\Gamma_{{\mathrm{total} ,j}}},
\label{eq.4}
\end{equation}
where $J_i$ is the target spin and $J_p=1/2$ is the proton spin. $J_j$, $\Gamma_{p,ij}$, $\Gamma_{\gamma, j}$, and $\Gamma_{\mathrm{total},j}$ are spin, proton decay width, $\gamma$-decay width, and total width of the compound nucleus state $j$, with $\Gamma_{\mathrm{total} ,j}  = \sum_{i}\Gamma_{p,ij} + \Gamma_{\gamma ,j}$.

For the calculation of the proton partial widths, the single-particle formalism~\cite{lane1960reduced,iliadis1997proton} is used, where the proton partial width for capture from the initial state $i$ through resonance state $j$ is given by
\begin{equation}
    \Gamma_{p,ij}
    =2P_l(E_{\mathrm{R},ij})\gamma^2_{p,ij}
    =2P_l(E_{\mathrm{R},ij})
    \frac{\hbar^2}{\mu R^2}
    \theta^2_{p,ij},
\label{eq.5}
\end{equation}
where $\gamma^2_{p,ij}$ and $\theta^2_{p,ij}$ are the reduced proton width and dimensionless reduced proton width, respectively. The channel radius is defined as $R = r_0(A_p^{1/3}+A_t^{1/3})$,
with $r_0=1.25$ fm, and $P_l(E_{\mathrm{R},ij})$ is the Coulomb penetrability for a given orbital angular momentum $l$ at the resonance energy $E_{\mathrm{R},ij}$.
For a single-particle state, the dimensionless reduced proton width is expressed as $\theta^2_{p,ij}=C^2S_{ij}\theta^2_{\mathrm{sp}}$,
where $C$ is the isospin Clebsch--Gordan coefficient, $S_{ij}$ is the nucleon spectroscopic factor for the transition from the initial state $i$ to the resonance state $j$, and $\theta^2_{\mathrm{sp}}$ is the dimensionless single-particle reduced width.
In the present work, the $\theta^2_{\rm sp}$ values are taken from \citet{9zv2-wlkl}, where they were determined using the formalism of~\citet{iliadis1997proton}.

\begin{figure}
\centering
\includegraphics[width=0.48\textwidth]{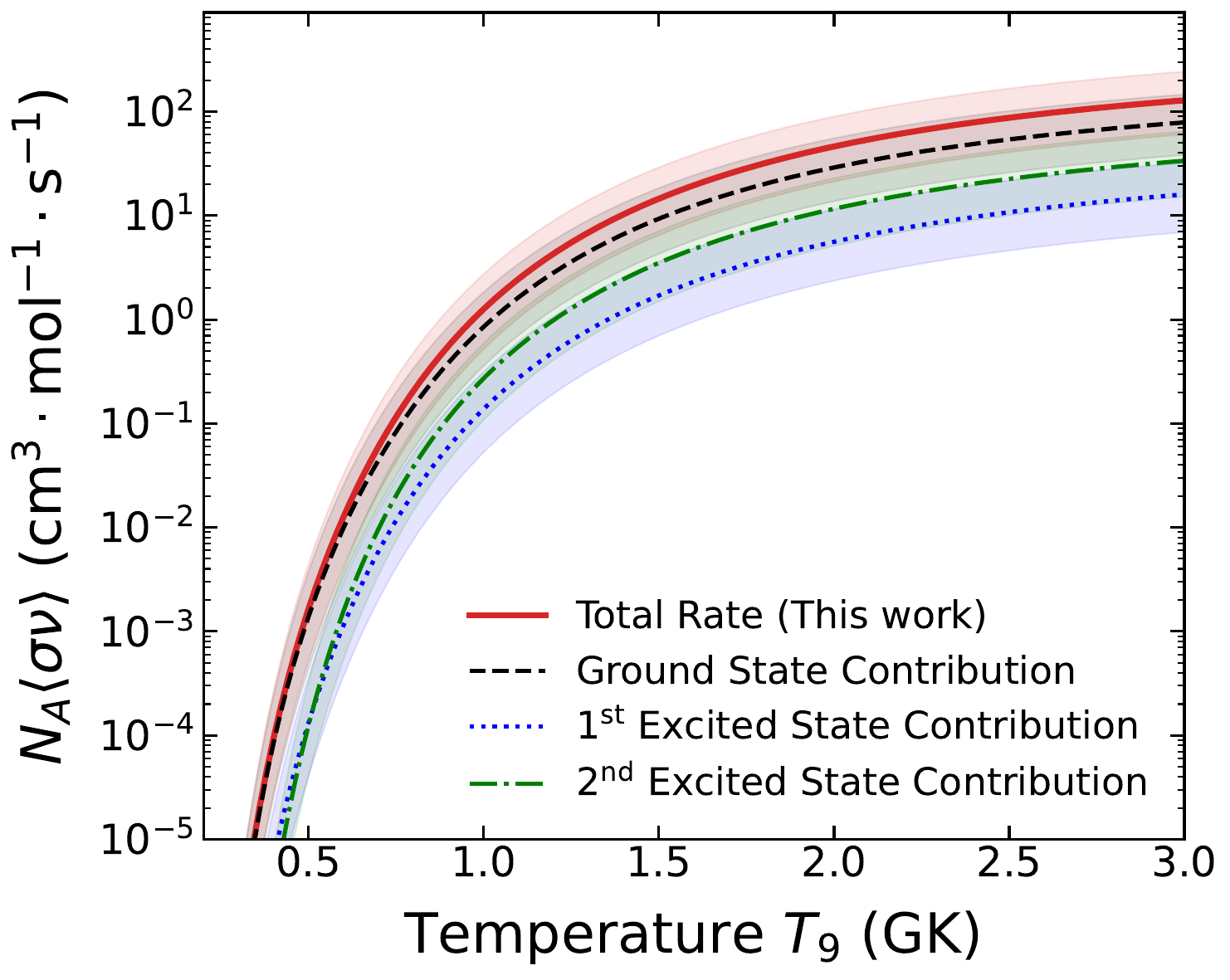}
\caption{The revised $^{45}$V($p,\gamma$)$^{46}$Cr reaction rate obtained in the present work, together with the individual contributions from proton capture on the ground state ($E=0$ keV, $J^\pi=7/2^-$), first excited state, ($E=56.7(5)$ keV, $J^\pi=5/2^-$) and second excited state ($E=56.8(6)$ keV, $J^\pi=3/2^-$) of $^{45}$V.}
\label{fig:rate1}
\end{figure}

\begin{figure}
\centering
\includegraphics[width=0.48\textwidth]{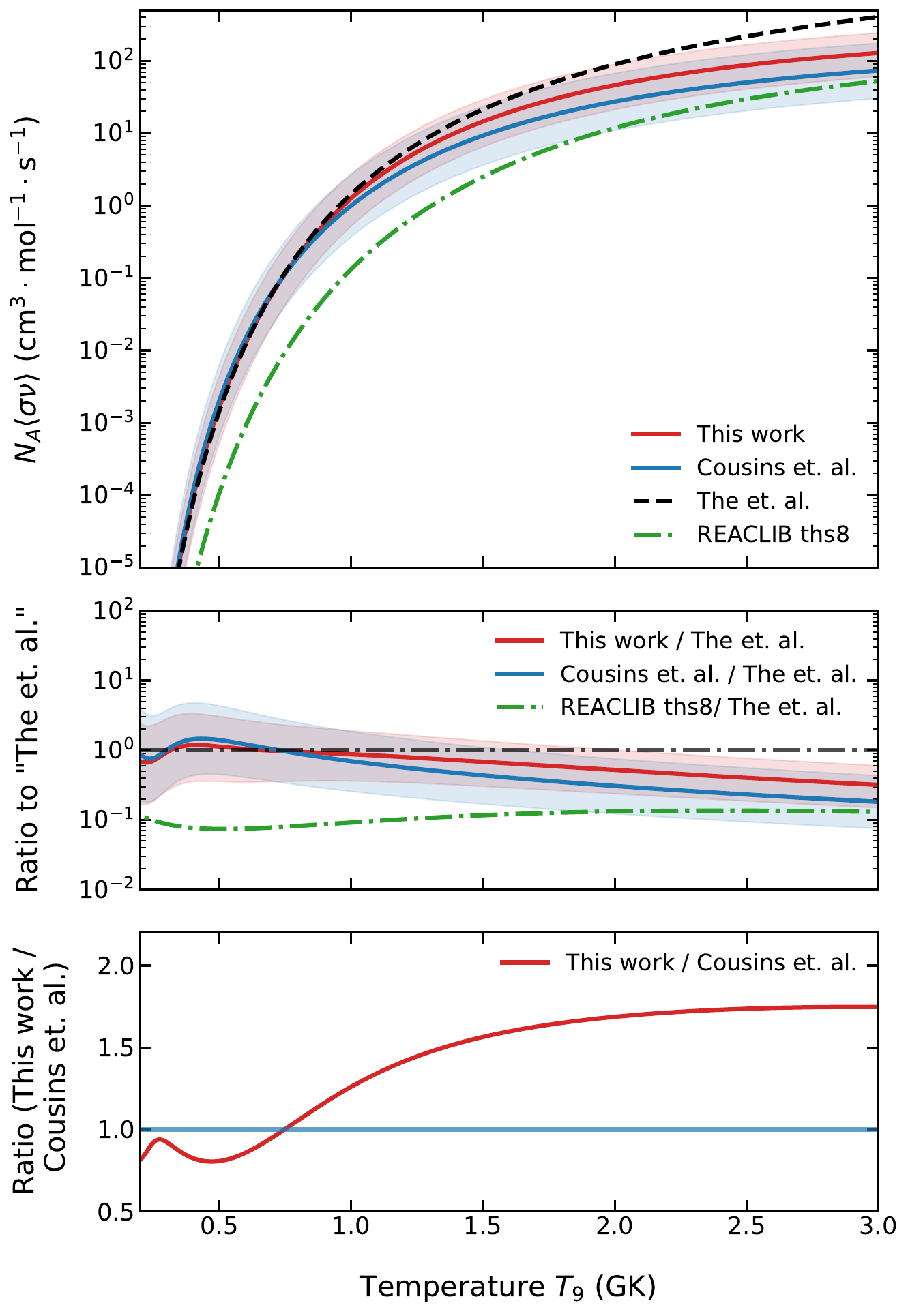}
\caption{Top: Comparison of the revised $^{45}$V($p,\gamma$)$^{46}$Cr reaction rate from the present work with those of Cousins \textit{et al.}~\cite{9zv2-wlkl}, The \textit{et al.}~\cite{The1998}, and REACLIB ths8~\cite{cyburt2010jina}. Middle: Ratios of the present work, Cousins \textit{et al.}~\cite{9zv2-wlkl}, and REACLIB ths8~\cite{cyburt2010jina} reaction rates relative to those of The \textit{et al.}~\cite{The1998}. Bottom: Ratio (without uncertainty bands) of the revised reaction rate from the present work to that of Cousins \textit{et al.}~\cite{9zv2-wlkl}. The deviation from unity remains within the combined uncertainties of the two reaction rates over the entire temperature range.}
\label{fig:rate2}
\end{figure}

To calculate the proton spectroscopic factors $C^2S$, 
shell-model calculations were performed using the KShell code~\cite{shimizu_thick-restart_2019} with the GXPF1A~\cite{gxpf1a} and SDPF-MU~\cite{sdpf_mu} effective interactions.
The GXPF1A interaction involves the $pf$ model space above the $^{40}$Ca core, whereas the SDPF-MU interaction involves the $sd$--$pf$ model space above the $^{16}$O core and allows cross-shell excitations.
As GXPF1A includes only the $pf$-shell orbitals, it is limited to describing negative-parity states in $^{45}$V and positive-parity states in $^{46}$Cr. On the other hand, SDPF-MU can describe both positive- and negative-parity states in $^{45}$V and $^{46}$Cr with suitable particle-hole ($p-h$) excitations.
The first three low-lying states of $^{45}$V, the ground state ($7/2^-$), first excited state ($5/2^-$), and second excited state ($3/2^-$), were calculated using both interactions. The calculated excitation energy of the first excited state ($5/2^-$) is 11 and 23 keV for GXPF1A and SDPF-MU, respectively, compared with the experimental value of 56.7(5) keV~\cite{BURROWS2008171}. For the second excited state ($3/2^-$), the calculated excitation energies are 350 and 321 keV using GXPF1A and SDPF-MU, respectively, which are higher than the experimental value of 56.8(6) keV~\cite{BURROWS2008171}. In the GXPF1A calculations, the full $pf$ shell was included, while the SDPF-MU calculations allowed $2p-2h$ excitations from the $sd$ shell.
For $^{46}$Cr, positive-parity states were calculated using both interactions within the $pf$ model space, and an average of the two results was taken. 
Negative-parity states were only calculated using the SDPF-MU interaction. 
The resulting $C^2S$ values are listed in Table~\ref{tab:table1} \footnote{It is to be noted that the $C^2S$ values for $^{45\mathrm{g}}$V($p,\gamma$)$^{46}$Cr obtained in the present shell-model calculations differ from those used by \citet{9zv2-wlkl}; the details are discussed in~\cite{Sidhu_supplement}.}. 
Following the approach of Cousins \textit{et al.}, a 40\% uncertainty was assigned to the $C^2S$ values.

The Coulomb penetrability is calculated using

\begin{equation}
    P_l(E_{\mathrm{R},ij}) = \frac{kR}{F_l^2(E_{\mathrm{R},ij}) + G_l^2(E_{\mathrm{R},ij})},
\label{eq.6}
\end{equation}
where $k$ is the wavenumber, $R$ is the channel radius, and $F_l$ and $G_l$ are the standard Coulomb functions, which can be calculated numerically.

The $\gamma$-ray widths $\Gamma_\gamma$ are adopted unchanged from~\citet{9zv2-wlkl}. Following the treatment of Cousins~\textit{et al.}, a factor of 1.7~\cite{iliadis1999explosive} uncertainty was assigned to the proton partial widths $\Gamma_p$, and a factor of 1.5 uncertainty was assigned to the $\gamma$-ray widths $\Gamma_\gamma$. Using Eqs.~\ref{eq.4}--\ref{eq.6}, the resonance strengths were calculated using all ten proton-unbound states. The results are listed in Table~\ref{tab:table1}, while the complete calculations, including uncertainties, are provided in Section~I and summarised in Tables~I--III in~\cite{Sidhu_supplement}.

The non-resonant direct proton capture contribution for the target nucleus in its initial state $i$ is parameterised in terms of the astrophysical $S$ factor and is calculated using~\cite{Fowler1967}

\begin{multline}
N_A  \langle\sigma v\rangle_{\mathrm{NR},i}\left[\frac{\rm{cm^3}}{\rm s\cdot mol}\right] =  7.8327 \times 10^{9}\left( \frac{Z}{\mu T_9^2}\right)^{1/3} \\ \times S_{\mathrm{eff},i}(E_0)  \times \mathrm{exp}\left[ -4.2487 \left( \frac{ Z^2 \mu}{T_9}\right)^{1/3} \right], 
\label{eq.7}
\end{multline}
where $Z$ = 23 is the charge number of the target nucleus and $S_{\mathrm{eff},i}(E_0)$ is the effective astrophysical $S$ factor at the Gamow peak energy $E_0$. The effective $S$ factor can be parameterised as
\begin{equation}
    \label{eq.8}
    S_{\mathrm{eff},i}(E_0) \approx S(0)_i \left[ 1 + 0.09807 \left( \frac{T_9}{Z_p^2Z_T^2 \mu} \right)^{1/3} \right],
\end{equation}
where $S_i(0)$ is the astrophysical $S$ factor at zero energy for proton capture on the target nucleus in its initial state $i$, expressed in units of MeV b.
The astrophysical $S$ factors for direct capture to the ground state of $^{46}$Cr have been calculated using the \texttt{RADCAP} code~\cite{bertulani2003radcap}. The spectroscopic factors adopted in the direct capture calculations are taken from the shell-model calculations. The resulting $S_i(0)$ values are $1.415\times10^{1}$, $5.788\times10^{-1}$, and $4.109\times10^{1}$~keV b for proton capture from the ground state, first excited state, and second excited state of $^{45}$V, respectively. Using Eqs.~\ref{eq.7} and~\ref{eq.8}, the direct capture contribution is calculated and found to be negligible (contributing less than 1\% of the total reaction rate at $T_9=2$). 
Therefore, the DC contribution is neglected in the final reaction rate calculations in the present work.

The total thermonuclear reaction rate was calculated using Eqs.~\ref{eq.3} and \ref{eq.4}, and the resulting rate is shown in Fig.~\ref{fig:rate1}. The red solid line represents the total revised $^{45}$V($p,\gamma$)$^{46}$Cr reaction rate obtained in the present work. The individual contributions from proton capture on the ground, first excited, and second excited states of $^{45}$V are shown as black (dashed), blue (dotted), and green (dash-dotted) lines, respectively. The total reaction-rate values are provided in Table~V in~\cite{Sidhu_supplement}.
Over the temperature range relevant to $\alpha$-rich freeze-out ($T_9 \simeq$ 1.5--2), proton capture on the ground state of $^{45}$V provides the dominant contribution to the total reaction rate. However, captures from the excited states are also significant, particularly from the second excited state. For example, at $T_9$ = 2, the contributions from the ground state, first excited state, and second excited state account for 62.7\%, 12.1\%, and 25.2\% of the total reaction rate, respectively.

The revised reaction rate from the present work is compared with those of Cousins \textit{et al.}~\cite{9zv2-wlkl}, The \textit{et al.}~\cite{The1998}, and REACLIBv2.2 ths8 (hereafter
referred to as REACLIB)~\cite{cyburt2010jina}.
The comparison is shown in Fig.~\ref{fig:rate2}.
The present rate, together with the rates of Cousins \textit{et al.} and REACLIB, is lower than the rate of The \textit{et al.} Compared with the rate of Cousins \textit{et al.}, the revised reaction rate is higher by up to 69\% at 2 GK. 
This increase results from the combined effects of the improved $^{46}$Cr mass, the inclusion of proton capture on the first and second excited states of $^{45}$V, and the newly calculated shell-model $C^2S$ values, with the latter providing the dominant contribution to the enhanced reaction rate.

The use of the improved $^{46}$Cr mass in the present work also addresses one of the dominant uncertainties identified by \citet{9zv2-wlkl}, namely the uncertainty in the proton separation energy, $S_p$, and consequently in the resonance energies, for which a more precise mass measurement of $^{46}$Cr was suggested to be important.
The present reaction-rate evaluation, which includes a fourfold improvement in the precision of the $^{46}$Cr mass~\cite{wang2022b}, demonstrates that the current mass precision is sufficient for reliable astrophysical rate calculations. Further improvements in the $^{46}$Cr mass precision would therefore have only a negligible impact on the total reaction rate. Future reductions in the reaction-rate uncertainty will instead require improved constraints on the remaining nuclear-structure inputs: $C^2S$ and $\Gamma_\gamma$, which dominate the overall uncertainty.

\section{Nucleosynthesis calculations}
\label{Section-III}

The nucleosynthesis of $^{44}\text{Ti}$ during CCSN is strongly dependent on the proton capture channel $^{44}\text{Ti}(p,\gamma)^{45}\text{V}$. 
Once $^{45}\text{V}$ is produced, the flow can either return to $^{44}\text{Ti}$ through photodisintegration, $^{45}$V($\gamma,p$)$^{44}$Ti, or move further up the proton-rich side via $^{45}\text{V}(p,\gamma)^{46}\text{Cr}$. 
In this work, we first use power-law parameterized trajectories to investigate the effect of the  $^{45}\text{V}(p,\gamma)^{46}\text{Cr}$ reaction starting from NSE conditions. The thermodynamic evolution follows the homologous-expansion form as used in previous sensitivity studies~\cite{Magkotsios2010,Subedi2020}
\begin{equation}
\rho(t) = \frac{\rho_0}{(2t+1)^3}, \quad T(t) = \frac{T_0}{2t+1},
\end{equation}
where $t$ is the time after the onset of the expansion, and $T_0$ and $\rho_0$ are the initial peak temperature and density, respectively.

\begin{figure}
\centering
\includegraphics[width=0.48\textwidth]{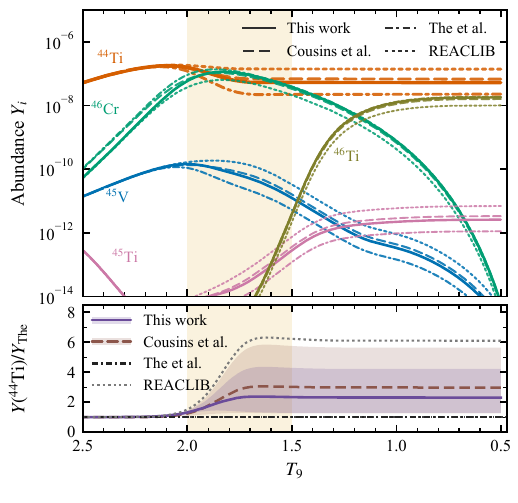}
\caption{The upper panel shows abundance evolution of key nuclei near $^{44}\text{Ti}$ as a function of $T_9$ for a representative power-law trajectory. The present revised, \citet{9zv2-wlkl}, The \textit{et al.}~\cite{The1998}, and REACLIB rates~\cite{Rauscher2002} are distinguished by solid, dashed, dash-dotted, and dotted lines, respectively. The lower panel shows the $Y(^{44}\mathrm{Ti})/Y_{\rm The}$ ratio for different rates. Shaded bands show the rate uncertainties for the present and Cousins \textit{et al.} rates.}
\label{fig:abun_evo}
\end{figure}

Figure~\ref{fig:abun_evo} illustrates the evolution of key abundances ($Y_i$) along a representative $\alpha$-rich freeze-out trajectory ($T_0 = 8.8$ GK, $\rho_0 = 1.58 \times 10^8\ \mathrm{g\ cm}^{-3}$, $Y_e = 0.50$), using the revised rate from present work, the Cousins \textit{et al.} rate~\cite{9zv2-wlkl}, the The \textit{et al.} rate~\cite{The1998}, and the REACLIB rate~\cite{Rauscher2002}. At high temperatures ($T_9 \gtrsim 5$), the composition is controlled by NSE, and the influence of individual proton-capture rates is negligible. 
As the temperature decreases through the $\alpha$-rich freeze-out window ($T_9 \simeq$ 2.0--1.5, shown by the yellow shaded region), the sequence $^{44}$Ti($p,\gamma$)$^{45}$V($p,\gamma$)$^{46}$Cr becomes active, allowing material to leak from the $^{44}$Ti--$^{45}$V quasi-equilibrium cluster.
The comparatively large rate of The \textit{et al.} drives the strongest leakage to $^{46}$Cr and consequently produces the lowest final $^{44}$Ti abundance. The revised reaction rate is higher than the central value reported by Cousins \textit{et al.} over this temperature range, resulting in greater leakage and a correspondingly lower final $^{44}$Ti abundance. At the same time, the smaller uncertainty of the revised rate leads to a narrower range of predicted final $^{44}$Ti abundances. Nevertheless, both experimentally constrained rates predict substantially higher $^{44}$Ti abundances than the rate of The \textit{et al.}, while the REACLIB rate yields the highest final $^{44}$Ti abundance. Once the temperature falls below $T_9\approx1.5$, proton captures are strongly suppressed by the Coulomb barrier and the $^{44}$Ti abundance freezes out.

\begin{figure*}
\centering
\includegraphics[width=0.85\textwidth]{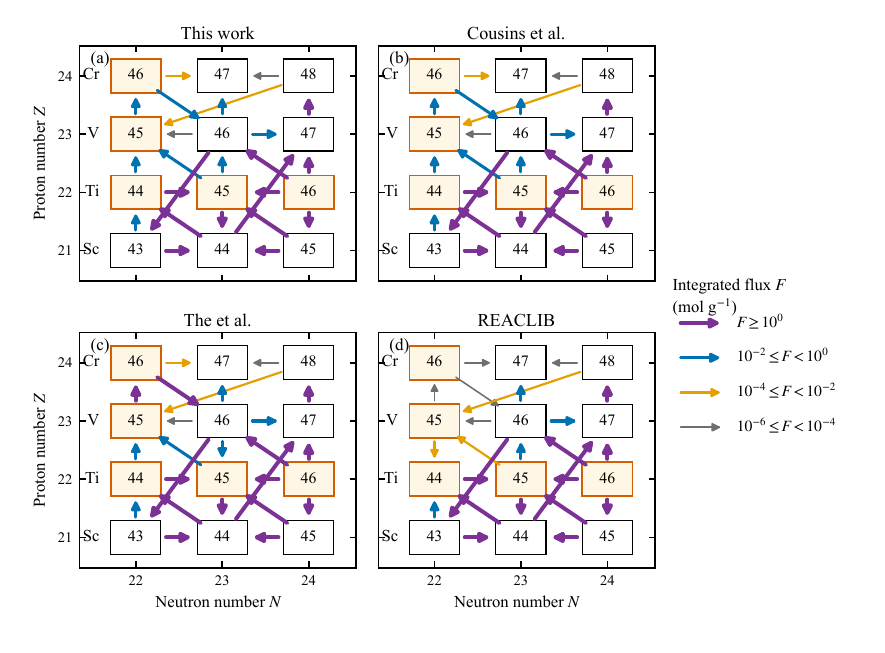}
\caption{The time-integrated net reaction flows among the neighboring Sc--Cr isotopes. Panel (a) to (d) represent the flow using the revised $^{45}$V($p,\gamma$)$^{46}$Cr rate from this work, the Cousins \textit{et al.} rate, the The \textit{et al.} rate, and the REACLIB rate, respectively.}
\label{fig:flow}
\end{figure*}

Figure~\ref{fig:flow} shows the time-integrated net reaction flows among the neighboring Sc--Cr isotopes. The overall flow pattern remains similar for the four adopted rates, while the most pronounced difference appears in the $^{45}$V$(p,\gamma)^{46}$Cr link. The rate of The \textit{et al.} produces the strongest integrated flow through this channel, enhancing the net flow in $^{44}$Ti--$^{45}$V region, leading to the additional flow from $^{46}$Cr to $^{46}$V and yielding the lowest final $^{44}$Ti abundance. The corresponding flow is weaker with the revised and Cousins \textit{et al.} rates, allowing more material to remain in the $^{44}$Ti region during freeze-out.

To study the impact of the revised $^{45}$V($p,\gamma$)$^{46}$Cr rate, we perform a $51 \times 51$ grid of nucleosynthesis calculations spanning peak temperatures $T_0 \in [4.0, 10.0]$ GK and peak densities $\log_{10}(\rho_0\ [\mathrm{g\ cm}^{-3}]) \in [4.0, 10.0]$ for two initial electron fractions: $Y_e = 0.50$ (symmetric matter) and $Y_e = 0.496$ (neutron-rich matter). The initial composition consists of $X(^{28}\text{Si}) = 1 - |2Y_e - 1|$ and a free neutron mass fraction $X(n) = |2Y_e - 1|$ for the neutron-rich case.

\begin{figure}
\centering
\includegraphics[width=0.49\textwidth]{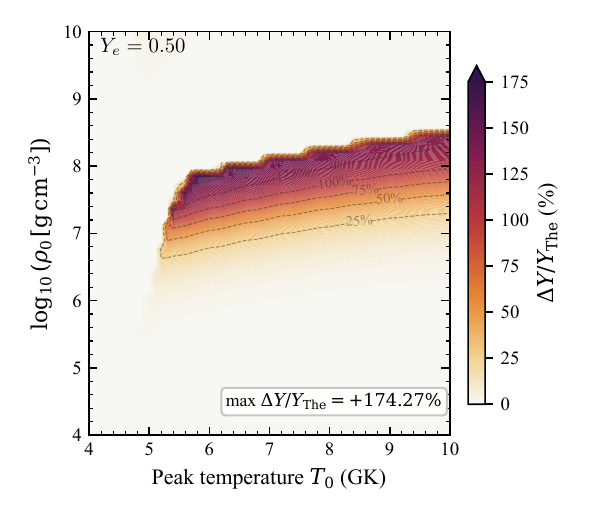}%
\caption{Relative change of the final $^{44}\text{Ti}$ yield produced by replacing the The \textit{et al.} $^{45}\text{V}(p,\gamma)^{46}\text{Cr}$ rate with the revised rate from present work in power-law trajectories. The main panel shows the $Y_e=0.50$ grid.}
\label{fig:grid_ye05}
\end{figure}

Under neutral conditions ($Y_e = 0.50$), Fig.~\ref{fig:grid_ye05} shows the relative change $\Delta Y/Y_{\rm ref}=(Y_{\rm present}-Y_{\rm ref})/Y_{\rm ref}$ in the final $^{44}\mathrm{Ti}$ abundance obtained with the present rate with respect to the The \textit{et al.} rate. Corresponding sensitivity maps relative to the rates of Cousins \textit{et al.} and REACLIB are shown in Fig. \ref{fig:grid_ye05_appen} of the Appendix. 
The largest sensitivity occurs for peak temperatures high enough to establish the QSE flow and for intermediate densities that maintain non-negligible proton abundance during freeze-out. In this region, the revised reaction rate reduces the final $^{44}$Ti yield by up to a factor of 1.7 relative to the rate of The \textit{et al.} At lower peak densities, proton captures are too slow to compete effectively before freeze-out. For the neutron-rich case ($Y_e=0.496$), however, the free-proton abundance during $\alpha$-rich freeze-out is too low to sustain significant flow through the $^{44}$Ti$(p,\gamma)^{45}$V$(p,\gamma)^{46}$Cr sequence, thereby suppressing leakage from the $^{44}$Ti region. Consequently, variations in the $^{45}$V($p,\gamma$)$^{46}$Cr reaction rate have only a negligible effect on the final $^{44}$Ti abundance, which only gives a $0.09\%$ change over the same grid (see Fig. \ref{fig:grid_ye0496_appen} of Appendix). 

\begin{figure}
\centering
\includegraphics[width=0.48\textwidth]{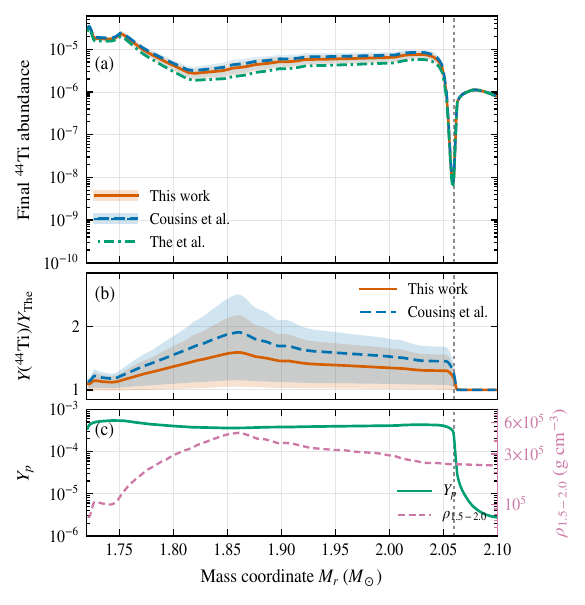}
\caption{Post-processing nucleosynthesis results for the
$20\,M_\odot$ CCSN trajectories as a function of mass coordinate
$M_r$. Panel (a) compares the final $^{44}$Ti abundances obtained
using the present, Cousins \textit{et al.}~\cite{9zv2-wlkl}, and The \textit{et al.}~\cite{The1998} rates. The shaded bands represent the propagated
rate uncertainties for the present and Cousins \textit{et al.} rates. Panel
(b) shows the corresponding abundance ratios relative to the The \textit{et al.} result. Panel (c) shows the maximum free-proton abundance $Y_p$
and density reached within the $T_9=1.5$--2.0 freeze-out window. The
vertical dotted line marks the outer boundary of the rate-sensitive
region near $M_r\simeq2.06\,M_\odot$.}
\label{fig:mr_profile}
\end{figure}

We further perform post-processing nucleosynthesis calculations using thermodynamic trajectories extracted from a 20~${M}_\odot$ CCSN model calculated by Modules for Experiments in Stellar Astrophysics (\texttt{MESA}) \cite{2019ApJS..243...10P} using a thermal bomb prescription. We inject energy into a thin shell of 0.05 $M_\odot$ for 20 ms with total energy $10^{51}$ erg. 
The mass cut is chosen at an entropy of $S=4k_B\rm\ baryon^{-1}$ \cite {Heger:2008td}. The post-processing nucleosynthesis calculations are then performed for the ejecta outside the mass cut.
Figure~\ref{fig:mr_profile} shows the final $^{44}$Ti abundance profile as a function of mass coordinate $M_r$. 
Both the present and Cousins \textit{et al.} rates produce higher $^{44}$Ti abundances than the The \textit{et al.} rate throughout most of the inner ejecta. The largest rate dependence occurs near $M_r\simeq1.86\,M_\odot$, where the local $^{44}$Ti abundance obtained with the present and Cousins \textit{et al.} rates is enhanced by factors of approximately 1.59 and 1.91, respectively, relative to the The \textit{et al.} rate. 
The lower abundance obtained with the present rate compared with Cousins \textit{et al.} is due to its larger $^{45}$V($p,\gamma$)$^{46}$Cr rate in the relevant freeze-out temperature range, which leads to stronger leakage from the $^{44}$Ti--$^{45}$V quasi-equilibrium cluster. 
The lower panel shows that the sensitive layers retain a free-proton abundance of several $10^{-4}$ during the $T_9=1.5$--2.0 freeze-out phase. The sensitivity reaches its maximum near the peak of the freeze-out density and then gradually decreases toward larger mass coordinate. Beyond $M_r\simeq2.06\,M_\odot$, $Y_p$ drops by more than two orders of magnitude and the three abundance profiles converge, indicating that proton-capture leakage through $^{45}$V$(p,\gamma)^{46}$Cr is no longer effective. 
Integrating the abundance profiles over the ejecta gives
$M(^{44}\mathrm{Ti})=8.93\times10^{-5}\,M_\odot$ for the The \textit{et al.} rate, $1.25^{+0.19}_{-0.29}\times10^{-4}\,M_\odot$ for the Cousins \textit{et al.} rate, and $1.13^{+0.21}_{-0.23}\times10^{-4}\,M_\odot$ for the present rate.

\begin{figure}
\centering
\includegraphics[width=0.48\textwidth]{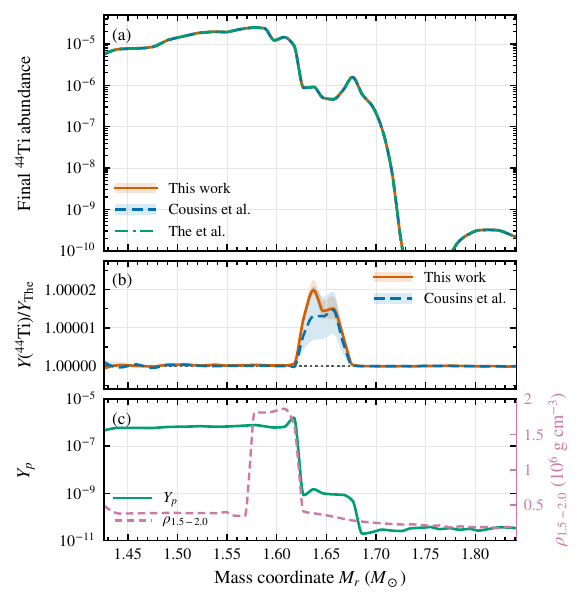}%
\caption{Post-processing nucleosynthesis results for the SN~1987A progenitor trajectories as a function of mass coordinate $M_r$. Panel (a) compares the final layer-by-layer $^{44}$Ti abundance obtained with the present, Cousins \textit{et al.}~\cite{9zv2-wlkl}, and The \textit{et al.}~\cite{The1998} rates; shaded bands show the propagated rate uncertainties for the present and Cousins \textit{et al.} rates. Panel (b) shows the corresponding $^{44}$Ti abundance ratios normalized to the The \textit{et al.} result. Panel (c) shows the maximum free-proton abundance $Y_p$ and maximum density reached while each trajectory passes through the rate-sensitive $T_9=1.5$--2.0 range.}
\label{fig:sn1987a_ti44}
\end{figure}

We also applied the same rate comparison to the SN~1987A model~\cite{1996ApJ...460..408T,2015PTEP.2015f3E01K}, with the mass cut defined by the ejection of $0.07\,M_\odot$ of $^{56}$Ni. In this case, all three reaction rates produce essentially identical ejected $^{44}$Ti masses of $M(^{44}\mathrm{Ti})\simeq1.31\times10^{-4}\,M_\odot$. As shown in Fig.~\ref{fig:sn1987a_ti44}, relative to the The \textit{et al.} rate, the present and Cousins \textit{et al.} rates change the integrated yield by only $3.9\times10^{-5}\%$ and $1.8\times10^{-5}\%$, respectively. Even in the most sensitive layer, near $M_r=1.637\,M_\odot$, the local abundance changes are only $2.0\times10^{-3}\%$ and $1.3\times10^{-3}\%$. These small differences remain stable when the integration tolerance is reduced to $10^{-8}$ and $10^{-9}$. In this model, the maximum free-proton abundance during the $T_9=1.5$--2.0 interval is only $Y_p\simeq1.6\times10^{-6}$, more than two orders of magnitude below that in the sensitive layers of the $20\,M_\odot$ model. Proton capture through $^{45}$V$(p,\gamma)^{46}$Cr is therefore strongly suppressed, making the final $^{44}$Ti yield effectively insensitive to the adopted rate. 

This contrasting behaviour explains the different conclusions reached by previous sensitivity studies. While The \textit{et al.}~\cite{The1998} identified $^{45}$V($p,\gamma$)$^{46}$Cr as an important reaction affecting $^{44}$Ti production, Subedi \textit{et al.}~\cite{Subedi2020} found little sensitivity of the integrated $^{44}$Ti yield. Our results indicate that the astrophysical impact of this reaction depends strongly on the $Y_e$ distribution of the $^{44}$Ti-producing ejecta. In proton-richer ejecta, such as the sensitive regions of the $20\,M_\odot$ model, sufficient free protons remain during freeze-out for the reaction to modify the final $^{44}$Ti abundance. In contrast, the slightly more neutron-rich SN~1987A ejecta have a lower proton abundance, suppressing this leakage pathway and making the final $^{44}$Ti yield insensitive to the reaction rate.

\begin{figure}
\centering
\includegraphics[width=0.48\textwidth]{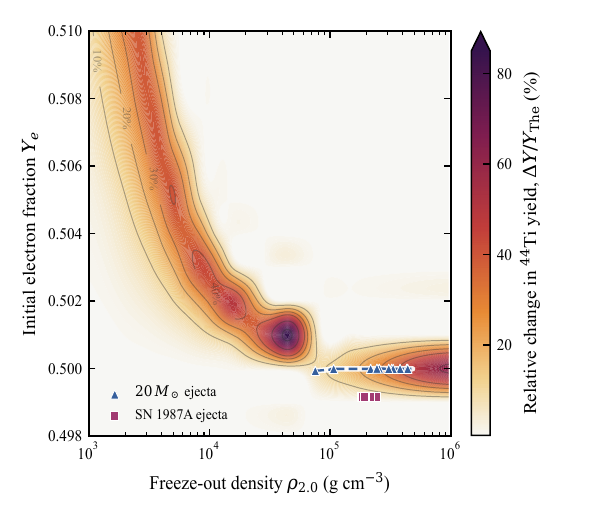}
\caption{Contours of the relative yield change of $^{44}\text{Ti}$ in the initial $Y_e$ and peak freeze-out density $\rho_{2.0}$ at $T_9 = 2.0$ plane.}
\label{fig:grid_ye_rho}
\end{figure}

To investigate this systematically, we perform a $21 \times 21$ grid sweep of initial electron fraction $Y_e \in [0.490, 0.510]$ and peak freeze-out density $\rho_{2.0} \in [10^3, 10^6]\ \mathrm{g\ cm}^{-3}$ at $T_9 = 2.0$. Figure~\ref{fig:grid_ye_rho} shows the relative change in the final $^{44}$Ti yield when the revised $^{45}$V($p,\gamma$)$^{46}$Cr rate is used instead of the The \textit{et al.}~\cite{The1998} rate.

The strongest reduction occurs in a narrow region near $Y_e\simeq0.50$--0.505 and $\rho_{2.0}\sim10^4$--$10^5\ {\rm g\ cm^{-3}}$, where enough free protons remain during freeze-out for $^{45}$V($p,\gamma$)$^{46}$Cr to provide an efficient leakage path out of the $^{44}$Ti--$^{45}$V region. The $20\,M_\odot$ ejecta remain very close to $Y_e=0.50$ and extend across $\rho_{2.0}\simeq7.5\times10^4$--$4.8\times10^5\ {\rm g\,cm^{-3}}$, placing part of the trajectory near the rate-sensitive region. By contrast, the ejected SN~1987A layers locate around the slightly lower value $Y_e\simeq0.499$ and therefore lie below the region of strongest sensitivity. 

\section{Summary}

In summary, we have adopted the precise B$\rho$-IMS mass measurement of $^{46}$Cr reported by Wang \textit{et al.}~\cite{wang2022b}, which provides a fourfold improvement in precision over the AME2020 value, to refine the astrophysical $^{45}$V($p,\gamma$)$^{46}$Cr reaction rate. 
In addition to proton capture on the ground state of $^{45}$V, we include contributions from proton capture on the first and second excited states. We also performed new shell-model calculations using the K-Shell code with GXPF1A and SDPF-MU interactions to determine the spectroscopic factors for proton capture on the ground, first excited, and second excited states of $^{45}$V. The revised reaction rate is up to 69\% higher than that reported by Cousins \textit{et al.} over the temperature range relevant to $\alpha$-rich freeze-out. 
This increase results from the combined effects of the improved $^{46}$Cr mass, the inclusion of proton capture on the first and second excited states of $^{45}$V, and the newly calculated shell-model $C^2S$ values, with the latter providing the dominant contribution to the enhanced reaction rate.

Since the proton spectroscopic factors remain model dependent, with uncertainties of a factor of $\sim$1.7, and the $\gamma$-decay widths are adopted unchanged from~\cite{9zv2-wlkl}, with uncertainties of a factor of $\sim$1.5, the uncertainty associated with the $^{46}$Cr mass is no longer the dominant contribution to the reaction-rate uncertainty.

Further improvements in the reaction rate require better experimental constraints on the $C^2S$ values and $\Gamma_\gamma$ widths, for example through high-resolution spectroscopy of  $^{46}$Cr and its mirror nucleus $^{46}$Ti or, ultimately, through direct $^{45}$V($p,\gamma$)$^{46}$Cr measurements in inverse kinematics at next-generation radioactive-beam facilities. In addition, high-resolution $\gamma$-ray spectroscopy studies of $^{45}$V are encouraged to better resolve the first and second excited states.

\begin{figure}
\centering
\includegraphics[width=0.48\textwidth]{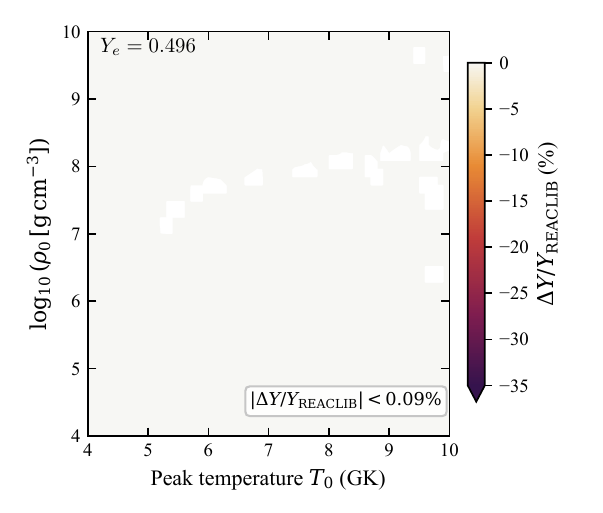}%
\caption{Relative change of the final $^{44}\text{Ti}$ yield produced by replacing the REACLIB rate with the revised rate in power-law trajectories. This panel show the $Y_e=0.496$ grid.}
\label{fig:grid_ye0496_appen}
\end{figure}
\begin{figure*}
\centering
\includegraphics[width=0.48\textwidth]{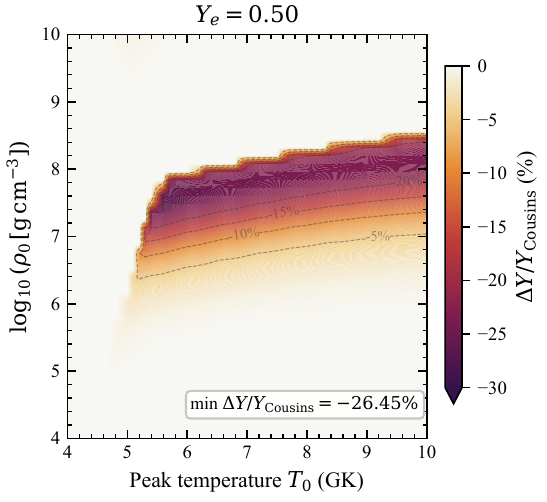}%
\includegraphics[width=0.48\textwidth]{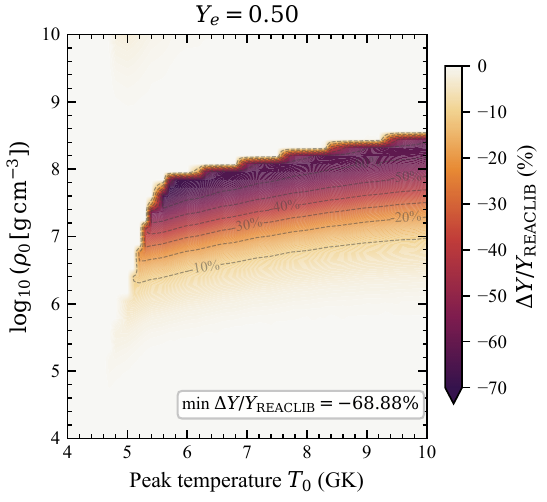}%
\caption{Relative change of the final $^{44}\text{Ti}$ yield produced by replacing the REACLIB $^{45}\text{V}(p,\gamma)^{46}\text{Cr}$ rate and Cousin \textit{et al.} rate with the revised rate in power-law trajectories. Both panel show the $Y_e=0.50$ grid.}
\label{fig:grid_ye05_appen}
\end{figure*}
We further performed detailed nucleosynthesis calculations using both parametrised thermodynamic trajectories and a $20\,M_\odot$ CCSN model generated with \texttt{MESA}, as well as an 
SN~1987A model. During the $\alpha$-rich freeze-out phase, the present $^{45}$V($p,\gamma$)$^{46}$Cr reaction rate is higher than the central rate reported by Cousins \textit{et al.}, leading to an increased reaction flow through $^{45}$V($p,\gamma$)$^{46}$Cr and consequently a lower final $^{44}$Ti abundance. For the $20\,M_\odot$ CCSN model, the present rate results in a final $^{44}$Ti yield of $M(^{44}\mathrm{Ti}) = 1.13^{+0.21}_{-0.23}\times10^{-4}\,M_\odot$, compared with $1.25^{+0.19}_{-0.29}\times10^{-4}\,M_\odot$ obtained using the Cousins \textit{et al.} rate and $8.93\times10^{-5}\,M_\odot$ using the The \textit{et al.} rate. 
In contrast, all three rate prescriptions give nearly identical $^{44}$Ti yields of approximately $1.31\times10^{-4}\,M_\odot$ for the SN~1987A model.
This comparison demonstrates that the sensitivity of the $^{45}$V($p,\gamma$)$^{46}$Cr reaction rate to the $^{44}$Ti yield depends strongly on the electron fraction ($Y_e$) of the ejecta. The reaction rate has a significant impact only in proton-rich ejecta with $Y_e$ close to 0.5, where the abundance of free protons is sufficiently high. At lower $Y_e$, the reduced proton abundance suppresses the reaction flow and therefore decreases the sensitivity of the $^{44}$Ti yield to the $^{45}$V($p,\gamma$)$^{46}$Cr reaction rate.
 These results help reconcile
the different conclusions of previous sensitivity studies
\citep{The1998,Magkotsios2010,Subedi2020} and demonstrate the
importance of the progenitor-dependent $Y_e$ distribution when
assessing the astrophysical impact of this reaction.

\appendix
\section{Impact of different reaction rates on the production of $^{44}$Ti}

Under neutron-rich conditions ($Y_e = 0.496$), the relative sensitivity to the reaction rate drops to zero as shown in Fig. \ref{fig:grid_ye0496_appen}.

Under neutral conditions ($Y_e = 0.50$), Fig.~\ref{fig:grid_ye05_appen} demonstrates the sensitivity of the results to the choice of the $^{45}\text{V}(p,\gamma)^{46}\text{Cr}$ reaction rates. In both comparisons, the sensitivity is concentrated in the high-density $\alpha$-rich freeze-out region, where proton captures efficiently divert material through $^{45}\mathrm{V}(p,\gamma)^{46}\mathrm{Cr}$. Relative to the~\citet{9zv2-wlkl} rate, the maximum reduction is $26.45\%$, occurring near $T_0=5.6$ GK and $\rho_0=3.98\times10^{7}\ {\rm g\,cm^{-3}}$ as shown in the left panel. In the right panel, the difference from the REACLIB rate is larger, reaching $68.88\%$ near $T_0=5.7$ GK and $\rho_0=5.25\times10^{7}\ {\rm g\,cm^{-3}}$. At lower densities, the proton-capture leakage becomes inefficient, and the calculated $^{44}$Ti abundance is nearly insensitive to the adopted rate.

\begin{acknowledgments}
The authors express their thanks to R. J. deBoer, C. Cousins, C. Iliadis, C. O'Shea, and C. Yuan for discussions.
This research was supported by the Science and Technologies Facilities Council (STFC), the National Science Foundation through Grant No.~PHY-2011890 (University of Notre Dame Nuclear Science Laboratory), Grant No.~PHY-1430152 (the Joint Institute for Nuclear Astrophysics - Center for the Evolution of the Elements), Grant No. OISE-1927130 (IReNA), the Youth Innovation Promotion Association of the Chinese Academy of Sciences (Grant No. 2022423), and the National Natural Science Foundation of China (Grant No 12322507).
\end{acknowledgments}

\end{document}


\title{Supplemental Material for: Revised $^{45}$V($p,\gamma$)$^{46}$Cr reaction rate and its impact on the production of $^{44}$Ti in core-collapse supernovae}

\author{R.~S.~Sidhu}
\email{ragan.sidhu@surrey.ac.uk}
\affiliation{School of Mathematics and Physics, University of Surrey, Guildford, GU2 7XH, United Kingdom}

\author{Y.~Luo}
\email{yudong.luo@pku.edu.cn}
\affiliation{School of Physics, and Kavli Institute for Astronomy and Astrophysics, Peking University, Beijing 100871, P. R. China, Beijing, Beijing, 100871, China}

\author{C.~Sarma}
\affiliation{School of Mathematics and Physics, University of Surrey, Guildford, GU2 7XH, United Kingdom}

\author{M.~Wiescher}
\affiliation{Department of Physics and Astronomy, University of Notre Dame, Notre Dame, Indiana 46556, USA}

\author{X.~Xu}
\affiliation{Institute of Modern Physics, Chinese Academy of Sciences, Lanzhou 730000, China}

\date{\today}

\pacs{}

\maketitle


\section{Resonance strength calculations}

The resonance strengths of the ten proton-unbound resonances were calculated using Eqs.~(4)--(6) of the main paper. The input parameters used in the calculations, together with the resulting resonance strengths for proton capture on the ground state, first excited state, and second excited state of $^{45}$V, are listed in Tables~\ref{tab:table1}--\ref{tab:table3}, respectively.

The resonance energies, $E_{\mathrm{R}}$, were determined using
$E_{\mathrm{R}} = E_x - S_p(^{46}$Cr) $-E_i$, 
where $E_x$ is the excitation energy of the resonant state in $^{46}$Cr, taken from Cousins \textit{et al.}~\cite{9zv2-wlkl}, and $E_i$ is the excitation energy of the initial state in $^{45}$V. The proton separation energy, $S_p(^{46}$Cr) = 4879.8(2.8) keV, was calculated using the mass excesses
ME($^{45}$V) = $-31886.4 \pm 0.9$ keV~\cite{wang2021ame} and ME($^{46}$Cr) = $-29477.2 \pm 2.6$ keV~\cite{wang2022b}.

The Coulomb penetrability, $P_l$, was calculated using Eq.~(6) of the main paper. To propagate the uncertainty arising from the resonance energy, penetrabilities were calculated at the central resonance energy ($E_R$), as well as at $E_R+\Delta E_R$ and $E_R-\Delta E_R$, and are reported in Tables~\ref{tab:table1}--\ref{tab:table3} as $P_l$ (median), $P_l$ (high), and $P_l$ (low), respectively\footnote{It is to be noted that in the study of \citet{9zv2-wlkl}, the penetrabilities were calculated at the $E_R$, and no uncertainty associated with the resonance energies was propagated to $P_l$.}.

The single-particle reduced widths, $\theta_{\rm sp}^2$
are taken from Cousins \textit{et al.}~\cite{9zv2-wlkl}, where they were determined using the formalism of~\citet{iliadis1997proton}. 
The spectroscopic factors, $C^2S$, for proton capture on the ground state, first excited state, and second excited state of $^{45}$V are calculated in the present work using the KShell code~\cite{shimizu_thick-restart_2019} with GXPF1A~\cite{gxpf1a} and SDPF-MU~\cite{sdpf_mu} effective shell-model interactions\footnote{The $C^2S$ values calculated using the shell model are given up to four decimal places.}.
Following their treatment, a 40\% uncertainty was assigned to the $C^2S$ values. 

It is to be noted that the $C^2S$ values for $^{45\mathrm{g}}$V($p,\gamma$)$^{46}$Cr obtained from the present shell-model calculations differ from those used by Cousins \textit{et al.}.
In that work, the $C^2S$ for the positive-parity states of $^{46}$Cr were calculated with a $0p$--$0h$ truncation for the $^{45}$V ground state, using the SDPF-MU interaction. On the other hand, for the negative parity states of $^{46}$Cr, the $^{45}$V ground state was calculated using a $2p$--$2h$ truncation. 
In the present work, all $C^2S$ values are calculated using a $2p$--$2h$ truncation with the SDPF-MU interaction for the three low-lying states of $^{45}$V.
For example, the calculated $6^+_4$ state in $^{46}$Cr, with excitation energies of 4999 and 4785 keV from the GXPF1A and SDPF-MU interactions, respectively, is assigned to the experimentally observed $6^+$ state at 4971(6) keV. The other shell-model states in $^{46}$Cr considered in the present calculation are listed in Table~\ref{tab:table1.0}. For each interaction, states in $^{46}$Cr with experimentally assigned $J^\pi$ values and excitation energies close to the measured levels are selected. For the three $4^+$ states considered in the reaction-rate calculation, we adopt the calculated $4^+_6$, $4^+_7$, and $4^+_{15}$ states, with excitation energies of 4972 (4724), 5722 (5601), and 7284 (7014) keV from the GXPF1A (SDPF-MU) interactions, respectively.
In cases where multiple shell-model states with the same $J^\pi$ values and similar excitation energies are available, the corresponding $J_i \rightarrow J_f$ transitions and $E_\gamma$ values reported in Table II of~\citet{9zv2-wlkl} are also considered to identify the most appropriate state assignments in $^{46}$Cr, with particular emphasis on the associated $E2$ and $M1$ transitions.
The good agreement between the calculated and experimental excitation energies, together with the matching spin-parity assignments, provides confidence in the correspondence between the shell-model and experimental states used to determine the $C^2S$ values in the present work.

The proton partial widths, $\Gamma_p$, were then calculated using Eq.~(5) of the main paper with the revised resonance energies and corresponding penetrabilities. In addition to the uncertainties arising from $P_l$ and $C^2S$, an overall uncertainty of a factor of 1.7~\cite{iliadis1999explosive} was applied to $\Gamma_p$, as done in the work of Cousins \textit{et al.}~\cite{9zv2-wlkl}.
The $\gamma$-ray partial widths, $\Gamma_\gamma$, were adopted directly from Cousins \textit{et al.}, together with their quoted uncertainty of a factor of 1.5. Finally, the resonance strengths, $\omega\gamma$, were calculated using Eq.~(4) of the main paper, and the corresponding uncertainties were obtained by propagating the uncertainties in the proton and $\gamma$-ray partial widths.
The resonance strengths, their associated uncertainties, and the parameters used in the resonance-strength calculations [Eqs.~(4)--(6) of the main text] are listed in Tables~\ref{tab:table1}--\ref{tab:table3} for proton capture on the ground state, first excited state, and second excited state of $^{45}$V, respectively.

\section{Thermonuclear reaction rate}

The total thermonuclear reaction rate \(N_A \langle\sigma v\rangle\) in the temperature range 0.1 \(< T_9 <\) 3  calculated using Eq.~(3) in the main paper, is provided in Table~\ref{tab:table4}.

\begin{table}[p] 
    \centering
    \begin{minipage}{\linewidth}
        \centering
        \setlength{\tabcolsep}{10pt} 
        \small                     
        
        \begin{threeparttable}
            \caption{Properties of the proton-unbound states in $^{46}$Cr for proton capture on the ground state of $^{45}$V. The values of $E_x$, $J^\pi$, $\theta^2_{\mathrm{sp}}$, and $\Gamma_\gamma$ are taken from Cousins \textit{et al.}~\cite{9zv2-wlkl}.}
            \label{tab:table1}
            
            \begin{tabular}{c c c c c c c c c c} 
                \toprule
                $E_x$ & $E_R$  & $J^\pi$ & $l_p$ & $\omega$ & $P_l$ & $P_l$ & $P_l$ & $\theta^2_{sp}$ & $C^2S$  \\
                (keV) &  (keV) &  &  &  & (median)  & (high)  & (low) &  &  \\
                \midrule
                4971(6)	&	91.2(6.6)	&	($6^+$)	&	3	&	0.81	&	1.62E-30	&	2.11E-29	&	9.30E-32	&	0.24	&	0.0184(74)	\\
5194(1)	&	314.2(2.9)	&	$2^+$	&	1	&	0.31	&	3.73E-13	&	4.50E-13	&	3.09E-13	&	0.53	&	0.1617(647)	\\
5305(5)	&	425.2(5.7)	&	($4^+$)	&	1	&	0.56	&	1.01E-10	&	1.27E-10	&	8.03E-11	&	0.53	&	0.0222(89)	\\
5411(1)	&	531.2(2.9)	&	($4^+$)	&	1	&	0.56	&	3.72E-09	&	4.06E-09	&	3.42E-09	&	0.53	&	0.2481(992)	\\
5588(2)	&	708.2(3.4)	&	$2^+$	&	1	&	0.31	&	2.21E-07	&	2.35E-07	&	2.07E-07	&	0.53	&	0.1077(431)	\\
5730(6)	&	850.2(6.6)	&	($5^+$)	&	1	&	0.69	&	2.19E-06	&	2.40E-06	&	1.99E-06	&	0.53	&	0.1802(721)	\\
5792(2)	&	912.2(3.4)	&	$2^+$	&	1	&	0.31	&	4.99E-06	&	5.21E-06	&	4.78E-06	&	0.53	&	0.0840(336)	\\
5829(6)	&	949.2(6.6)	&	($3^-$)	&	0	&	0.44	&	2.32E-05	&	2.51E-05	&	2.14E-05	&	0.40	&	0.0010(4)	\\
6665(5)	&	1785.2(5.7)	&	($2^-$)	&	2	&	0.31	&	5.41E-04	&	5.55E-04	&	5.27E-04	&	0.40	&	0.0001(0)	\\
7120(7)	&	2240.2(7.5)	&	($4^+$)	&	1	&	0.56	&	1.61E-02	&	1.65E-02	&	1.57E-02	&	0.53	&	0.0517(207)	\\

                \bottomrule
            \end{tabular}
        \end{threeparttable}
    \end{minipage}
    
    \vspace{1.5cm} 
    
    \begin{minipage}{\linewidth}
        \centering
        \setlength{\tabcolsep}{8pt} 
        \small               
        
        \begin{threeparttable}
            \begin{tabular}{c c c c c c c c c }
                \toprule
                $\Gamma_p$ (eV)  & $\Gamma_p$ (eV)  & $\Gamma_p$ (eV) & $\Gamma_\gamma$ (eV) & $\Gamma_\gamma$ (eV) & $\Gamma_\gamma$ (eV) & $\omega\gamma$ (eV) & $\omega\gamma$ (eV) & $\omega\gamma$ (eV)  \\
                (median)  & (high)  & (low) & (median)  & (high)  & (low) & (median)  & (high)  & (low)   \\
                \midrule
1.87E-26	&	5.79E-25	&	3.79E-28	&	9.51E-02	&	1.43E-01	&	6.34E-02	&	1.52E-26	&	4.71E-25	&	3.08E-28	\\
8.35E-08	&	2.40E-07	&	2.45E-08	&	3.49E-01	&	5.24E-01	&	2.33E-01	&	2.61E-08	&	7.51E-08	&	7.65E-09	\\
3.12E-06	&	9.31E-06	&	8.72E-07	&	4.81E-02	&	7.22E-02	&	3.21E-02	&	1.75E-06	&	5.24E-06	&	4.90E-07	\\
1.28E-03	&	3.32E-03	&	4.15E-04	&	4.98E-01	&	7.47E-01	&	3.32E-01	&	7.18E-04	&	1.86E-03	&	2.33E-04	\\
3.30E-02	&	8.36E-02	&	1.09E-02	&	7.22E-01	&	1.08E+00	&	4.81E-01	&	9.55E-03	&	2.30E-02	&	3.29E-03	\\
5.46E-01	&	1.43E+00	&	1.76E-01	&	9.21E-02	&	1.38E-01	&	6.14E-02	&	5.41E-02	&	8.65E-02	&	3.13E-02	\\
5.81E-01	&	1.44E+00	&	1.96E-01	&	8.75E-01	&	1.31E+00	&	5.83E-01	&	7.39E-02	&	1.32E-01	&	3.53E-02	\\
2.43E-02	&	6.24E-02	&	7.90E-03	&	3.16E-01	&	4.74E-01	&	2.11E-01	&	9.40E-03	&	2.23E-02	&	3.25E-03	\\
5.66E-02	&	1.38E-01	&	1.94E-02	&	3.69E-01	&	5.54E-01	&	2.46E-01	&	6.08E-03	&	1.05E-02	&	3.06E-03	\\
1.15E+03	&	2.81E+03	&	3.98E+02	&	8.12E-01	&	1.22E+00	&	5.41E-01	&	3.98E-01	&	5.96E-01	&	2.65E-01	\\

                \bottomrule
            \end{tabular}
        \end{threeparttable}
    \end{minipage}

\end{table}

\begin{table}[p] 
    \centering
    
    \begin{minipage}{\linewidth}
        \centering
        \setlength{\tabcolsep}{10pt} 
        \small                     
        
        \begin{threeparttable}
            \caption{Properties of the proton-unbound states in $^{46}$Cr for proton capture on the first excited state of $^{45}$V. The values of $E_x$, $J^\pi$, $\theta^2_{\mathrm{sp}}$, and $\Gamma_\gamma$ are taken from Cousins \textit{et al.}~\cite{9zv2-wlkl}.}
            \label{tab:table2}
            
            \begin{tabular}{c c c c c c c c c c} 
                \toprule
                $E_x$ & $E_R$  & $J^\pi$ & $l_p$ & $\omega$ & $P_l$ & $P_l$ & $P_l$ & $\theta^2_{sp}$ & $C^2S$  \\
                (keV) &  (keV) &  &  &  & (median)  & (high)  & (low) &  &  \\
                \midrule
4971(6)	&	34.5(66)	&	($6^+$)	&	3	&	1.08	&	8.49E-51	&	2.20E-46	&	1.07E-56	&	0.24	&	0.0328(131)	\\
5194(1)	&	257.5(30)	&	$2^+$	&	1	&	0.42	&	5.67E-15	&	7.32E-15	&	4.37E-15	&	0.53	&	0.1286(514)	\\
5305(5)	&	368.5(57)	&	($4^+$)	&	1	&	0.75	&	7.95E-12	&	1.05E-11	&	5.96E-12	&	0.53	&	0.0433(173)	\\
5411(1)	&	474.5(30)	&	($4^+$)	&	1	&	0.75	&	6.30E-10	&	6.98E-10	&	5.68E-10	&	0.53	&	0.0925(370)	\\
5588(2)	&	651.5(34)	&	$2^+$	&	1	&	0.42	&	7.19E-08	&	7.72E-08	&	6.69E-08	&	0.53	&	0.0788(315)	\\
5730(6)	&	793.5(66)	&	($5^+$)	&	3	&	0.92	&	6.46E-09	&	7.19E-09	&	5.79E-09	&	0.53	&	0.0565(226)	\\
5792(2)	&	855.5(34)	&	$2^+$	&	1	&	0.42	&	2.35E-06	&	2.47E-06	&	2.25E-06	&	0.53	&	0.0606(242)	\\
5829(6)	&	892.5(66)	&	($3^-$)	&	0	&	0.58	&	1.15E-05	&	1.25E-05	&	1.06E-05	&	0.40	&	0.0013(5)	\\
6665(5)	&	1728.5(57)	&	($2^-$)	&	0	&	0.42	&	6.19E-03	&	6.35E-03	&	6.04E-03	&	0.40	&	0.0000(0)	\\
7120(7)	&	2183.5(75)	&	($4^+$)	&	1	&	0.75	&	1.36E-02	&	1.39E-02	&	1.32E-02	&	0.53	&	0.0090(36)	\\

                \bottomrule
            \end{tabular}
        \end{threeparttable}
    \end{minipage}
    
    \vspace{1.5cm} 
    
    \begin{minipage}{\linewidth}
        \centering
        \setlength{\tabcolsep}{8pt} 
        \small               
        
        \begin{threeparttable}
            \begin{tabular}{c c c c c c c c c }
                \toprule
                $\Gamma_p$ (eV)  & $\Gamma_p$ (eV)  & $\Gamma_p$ (eV) & $\Gamma_\gamma$ (eV) & $\Gamma_\gamma$ (eV) & $\Gamma_\gamma$ (eV) & $\omega\gamma$ (eV) & $\omega\gamma$ (eV) & $\omega\gamma$ (eV)  \\
                (median)  & (high)  & (low) & (median)  & (high)  & (low) & (median)  & (high)  & (low)   \\
                \midrule
1.75E-46	&	1.08E-41	&	7.81E-53	&	9.51E-02	&	1.43E-01	&	6.34E-02	&	1.89E-46	&	1.17E-41	&	8.46E-53	\\
1.01E-09	&	3.11E-09	&	2.75E-10	&	3.49E-01	&	5.24E-01	&	2.33E-01	&	4.21E-10	&	1.29E-09	&	1.15E-10	\\
4.77E-07	&	1.51E-06	&	1.26E-07	&	4.81E-02	&	7.22E-02	&	3.21E-02	&	3.58E-07	&	1.13E-06	&	9.47E-08	\\
8.08E-05	&	2.13E-04	&	2.57E-05	&	4.98E-01	&	7.47E-01	&	3.32E-01	&	6.04E-05	&	1.59E-04	&	1.93E-05	\\
7.85E-03	&	2.01E-02	&	2.58E-03	&	7.22E-01	&	1.08E+00	&	4.81E-01	&	3.03E-03	&	7.38E-03	&	1.03E-03	\\
5.06E-04	&	1.34E-03	&	1.60E-04	&	9.21E-02	&	1.38E-01	&	6.14E-02	&	6.69E-05	&	1.09E-04	&	3.80E-05	\\
1.98E-01	&	4.93E-01	&	6.66E-02	&	8.75E-01	&	1.31E+00	&	5.83E-01	&	3.36E-02	&	6.02E-02	&	1.60E-02	\\
1.57E-02	&	4.06E-02	&	5.08E-03	&	3.16E-01	&	4.74E-01	&	2.11E-01	&	8.09E-03	&	1.94E-02	&	2.79E-03	\\
0.00E+00	&	0.00E+00	&	0.00E+00	&	3.69E-01	&	5.54E-01	&	2.46E-01	&	0.00E+00	&	0.00E+00	&	0.00E+00	\\
1.69E+02	&	4.12E+02	&	5.83E+01	&	8.12E-01	&	1.22E+00	&	5.41E-01	&	7.77E-02	&	1.17E-01	&	5.17E-02	\\

                \bottomrule
            \end{tabular}
        \end{threeparttable}
    \end{minipage}

\end{table}

\begin{table}[p] 
    \centering
    
    \begin{minipage}{\linewidth}
        \centering
        \setlength{\tabcolsep}{10pt} 
        \small                     
        
        \begin{threeparttable}
            \caption{Properties of the proton-unbound states in $^{46}$Cr for proton capture on the second excited state of $^{45}$V. The values of $E_x$, $J^\pi$, $\theta^2_{\mathrm{sp}}$, and $\Gamma_\gamma$ are taken from Cousins \textit{et al.}~\cite{9zv2-wlkl}.}
            \label{tab:table3}
            
            \begin{tabular}{c c c c c c c c c c} 
                \toprule
                $E_x$ & $E_R$  & $J^\pi$ & $l_p$ & $\omega$ & $P_l$ & $P_l$ & $P_l$ & $\theta^2_{sp}$ & $C^2S$  \\
                (keV) &  (keV) &  &  &  & (median)  & (high)  & (low) &  &  \\
                \midrule
4971(6)	&	34.4(66)	&	($6^+$)	&	5	&	1.63	&	7.39E-55	&	2.03E-50	&	8.58E-61	&	0.24	&	0.0000(0)	\\
5194(1)	&	257.4(30)	&	$2^+$	&	1	&	0.63	&	5.62E-15	&	7.26E-15	&	4.33E-15	&	0.53	&	0.0581(232)	\\
5305(5)	&	368.4(57)	&	($4^+$)	&	3	&	1.13	&	3.67E-14	&	4.90E-14	&	2.74E-14	&	0.53	&	0.0250(100)	\\
5411(1)	&	474.4(30)	&	($4^+$)	&	3	&	1.13	&	3.23E-12	&	3.58E-12	&	2.90E-12	&	0.53	&	0.0034(13)	\\
5588(2)	&	651.4(35)	&	$2^+$	&	1	&	0.63	&	7.17E-08	&	7.72E-08	&	6.66E-08	&	0.53	&	0.1609(643)	\\
5730(6)	&	793.4(66)	&	($5^+$)	&	3	&	1.38	&	6.45E-09	&	7.18E-09	&	5.78E-09	&	0.53	&	0.0010(4)	\\
5792(2)	&	855.4(35)	&	$2^+$	&	1	&	0.63	&	2.35E-06	&	2.47E-06	&	2.24E-06	&	0.53	&	0.1519(608)	\\
5829(6)	&	892.4(66)	&	($3^-$)	&	2	&	0.88	&	4.95E-07	&	5.41E-07	&	4.53E-07	&	0.40	&	0.0023(9)	\\
6665(5)	&	1728.4(57)	&	($2^-$)	&	0	&	0.63	&	6.19E-03	&	6.35E-03	&	6.04E-03	&	0.40	&	0.0001(0)	\\
7120(7)	&	2183.4(75)	&	($4^+$)	&	3	&	1.13	&	2.65E-04	&	2.72E-04	&	2.57E-04	&	0.53	&	0.0050(20)	\\

                \bottomrule
            \end{tabular}
        \end{threeparttable}
    \end{minipage}
    
    \vspace{1.5cm} 
    
    \begin{minipage}{\linewidth}
        \centering
        \setlength{\tabcolsep}{8pt} 
        \small               
        
        \begin{threeparttable}
            \begin{tabular}{c c c c c c c c c }
                \toprule
                $\Gamma_p$ (eV)  & $\Gamma_p$ (eV)  & $\Gamma_p$ (eV) & $\Gamma_\gamma$ (eV) & $\Gamma_\gamma$ (eV) & $\Gamma_\gamma$ (eV) & $\omega\gamma$ (eV) & $\omega\gamma$ (eV) & $\omega\gamma$ (eV)  \\
                (median)  & (high)  & (low) & (median)  & (high)  & (low) & (median)  & (high)  & (low)   \\
                \midrule
               0.00E+00	&	0.00E+00	&	0.00E+00	&	9.51E-02	&	1.43E-01	&	6.34E-02	&	0.00E+00	&	0.00E+00	&	0.00E+00	\\
4.53E-10	&	1.39E-09	&	1.23E-10	&	3.49E-01	&	5.24E-01	&	2.33E-01	&	2.83E-10	&	8.69E-10	&	7.70E-11	\\
1.27E-09	&	4.04E-09	&	3.35E-10	&	4.81E-02	&	7.22E-02	&	3.21E-02	&	1.43E-09	&	4.54E-09	&	3.77E-10	\\
1.52E-08	&	4.02E-08	&	4.83E-09	&	4.98E-01	&	7.47E-01	&	3.32E-01	&	1.71E-08	&	4.50E-08	&	5.43E-09	\\
1.60E-02	&	4.10E-02	&	5.24E-03	&	7.22E-01	&	1.08E+00	&	4.81E-01	&	9.27E-03	&	2.26E-02	&	3.15E-03	\\
8.94E-06	&	2.37E-05	&	2.83E-06	&	9.21E-02	&	1.38E-01	&	6.14E-02	&	1.77E-06	&	2.88E-06	&	1.01E-06	\\
4.95E-01	&	1.24E+00	&	1.66E-01	&	8.75E-01	&	1.31E+00	&	5.83E-01	&	1.26E-01	&	2.26E-01	&	5.99E-02	\\
1.19E-03	&	3.10E-03	&	3.85E-04	&	3.16E-01	&	4.74E-01	&	2.11E-01	&	9.23E-04	&	2.21E-03	&	3.17E-04	\\
6.48E-01	&	1.58E+00	&	2.23E-01	&	3.69E-01	&	5.54E-01	&	2.46E-01	&	1.39E-01	&	2.41E-01	&	7.02E-02	\\
1.83E+00	&	4.49E+00	&	6.29E-01	&	8.12E-01	&	1.22E+00	&	5.41E-01	&	1.26E-03	&	1.91E-03	&	8.37E-04	\\

                \bottomrule
            \end{tabular}
        \end{threeparttable}
    \end{minipage}

\end{table}

\begin{table*}
\begin{ruledtabular}
\caption{Experimental excitation energies ($E_x$) and spin-parity assignments of the proton-unbound states in $^{46}$Cr adopted from Cousins \textit{et al.}~\cite{9zv2-wlkl} are listed in Columns I and II. Columns III--VI give the corresponding shell-model excitation energies and spin-parity assignments obtained with the GXPF1A and SDPF-MU interactions. The shell-model states shown are those matched to the experimental states and used to determine the $C^2S$ values in the present work.} 
\begin{tabular}{ c c  c c  c c }                    
 Counsins \textit{et al.}~\cite{9zv2-wlkl} &  & GXPF1A &  & SDPF-MU &   \\
 $E_x$ (keV) & $J^\pi$ &$E_x$ (keV) & $J^\pi$ & $E_x$ (keV) & $J^\pi$  \\\hline  
4971(6)	&	$6^+$	&	4999 & 6$^+_4$	 &	4785	&	$6^+_4$	\\
5194(1)	&	$2^+$	&	5009 & $2^+_5$& 5214
& $2^+_6$ \\
5305(5)	&	$4^+$	&	4972 & $4^+_6$ & 4724
& $4^+_6$\\
5411(1)	&	$4^+$	&	5722 & $4^+_7$ & 5601
 &	$4^+_7$ \\
5588(2)	&	$2^+$	&	5535 & $2^+_7$ & 5614 & $2^+_8$	\\
5730(6)	&	$5^+$	&	5736
& $5^+_5$ & 5519
&	$5^+_5$ \\
5892(2)	&	$2^+$	&	5815 & $2^+_8$ & 5837
& $2^+_9$ \\
5929(6)	&	$3^-$	&	--- & --- & 6062 & $3^-_{11}$	\\
6665(5)	&	$2^-$	&	--- & --- & 5704 & $2^-_7$	\\
7120(7)	&	$4^+$	&	7284 & $4^+_{15}$& 7014 & $4^+_{15}$  \\

\end{tabular}
\label{tab:table1.0}
\end{ruledtabular}
\end{table*}

\begin{table*}
\begin{ruledtabular}
\caption{Total thermonuclear reaction rate $N_A \langle\sigma v\rangle$ for the 
$^{45}$V$p,\gamma$)$^{46}$Cr reaction obtained from the 10 resonances in units $\rm{cm^3 s^{-1} mol^{-1}}$, as a function of temperature
$T_9$ .} 
\begin{center}
\begin{tabular}{ l l l l }              
$T_9$  &     Median Rate        &    High Rate       &         Low Rate     \\ \hline

0.1	&	1.93E-17	&	7.80E-17	&	4.03E-18	\\
0.2	&	6.58E-10	&	2.28E-09	&	1.61E-10	\\
0.3	&	1.02E-06	&	3.10E-06	&	2.85E-07	\\
0.4	&	9.42E-05	&	2.67E-04	&	2.80E-05	\\
0.5	&	1.65E-03	&	4.46E-03	&	5.18E-04	\\
0.6	&	1.27E-02	&	3.26E-02	&	4.25E-03	\\
0.7	&	6.04E-02	&	1.47E-01	&	2.17E-02	\\
0.8	&	2.08E-01	&	4.83E-01	&	7.90E-02	\\
0.9	&	5.62E-01	&	1.26E+00	&	2.23E-01	\\
1.0	&	1.26E+00	&	2.74E+00	&	5.20E-01	\\
1.1	&	2.47E+00	&	5.23E+00	&	1.04E+00	\\
1.2	&	4.32E+00	&	8.97E+00	&	1.86E+00	\\
1.3	&	6.91E+00	&	1.41E+01	&	3.02E+00	\\
1.4	&	1.03E+01	&	2.09E+01	&	4.55E+00	\\
1.5	&	1.46E+01	&	2.91E+01	&	6.48E+00	\\
1.6	&	1.96E+01	&	3.88E+01	&	8.79E+00	\\
1.7	&	2.54E+01	&	4.99E+01	&	1.15E+01	\\
1.8	&	3.18E+01	&	6.22E+01	&	1.44E+01	\\
1.9	&	3.88E+01	&	7.56E+01	&	1.77E+01	\\
2.0	&	4.63E+01	&	8.97E+01	&	2.12E+01	\\
2.1	&	5.41E+01	&	1.05E+02	&	2.49E+01	\\
2.2	&	6.22E+01	&	1.20E+02	&	2.87E+01	\\
2.3	&	7.05E+01	&	1.35E+02	&	3.26E+01	\\
2.4	&	7.89E+01	&	1.51E+02	&	3.66E+01	\\
2.5	&	8.74E+01	&	1.67E+02	&	4.06E+01	\\
2.6	&	9.58E+01	&	1.82E+02	&	4.47E+01	\\
2.7	&	1.04E+02	&	1.98E+02	&	4.87E+01	\\
2.8	&	1.12E+02	&	2.13E+02	&	5.26E+01	\\
2.9	&	1.21E+02	&	2.28E+02	&	5.65E+01	\\
3.0	&	1.28E+02	&	2.43E+02	&	6.04E+01	\\

\end{tabular}
\label{tab:table4}
\end{center}
\end{ruledtabular}
\end{table*}